\documentclass[twocolumn,showpacs,floats,floatfix,superscriptaddress,aps,pra]{revtex4-1}

\usepackage{amsfonts}
\usepackage{amssymb}
\usepackage{amsmath}
\usepackage{calc}
\usepackage{dcolumn}
\usepackage{graphicx}
\usepackage{bm}
\usepackage{epstopdf}
\usepackage{epsfig}
\usepackage{float}
\usepackage {color}
\usepackage{textcomp}
\usepackage{soul}
\usepackage{diagbox}

\begin{document}

\title{Polarization Independent Optical Isolator in Sagnac-Type Configuration}

\author{Mouhamad Al-Mahmoud}
\affiliation{Department of Physics, Sofia University, James Bourchier 5 blvd, 1164 Sofia, Bulgaria}
\email{mouhamadmahmoud1@gmail.com}

\author{Emiliya Dimova}
\affiliation{Institute of Solid State Physics, Bulgarian Academy of Sciences, 72 Tsarigradsko chauss\'{e}e, 1784 Sofia, Bulgaria}

\author{Hristina Hristova}
\affiliation{Institute of Solid State Physics, Bulgarian Academy of Sciences, 72 Tsarigradsko chauss\'{e}e, 1784 Sofia, Bulgaria}

\author{Virginie Coda}
\affiliation{Universit\'e de Lorraine, CentraleSup\'elec, LMOPS, F-57000 Metz, France}

\author{Andon Rangelov}
\affiliation{Department of Physics, Sofia University, James Bourchier 5 blvd, 1164 Sofia, Bulgaria}

\author{Germano Montemezzani}
\affiliation{Universit\'e de Lorraine, CentraleSup\'elec, LMOPS, F-57000 Metz, France}

\begin{abstract}
We propose a polarization-independent optical isolator that does not use walk-off between the two orthogonal polarizations. The design is based on two Faraday rotators in combination with two half-wave plates in a closed, Sagnac-interferometer-like, configuration. An experimental prototype is tested successfully under variation of the input light polarization state with isolation level between 43 dB and 50 dB for all input polarizations (linear, circular, or elliptical).
\end{abstract}

\maketitle

\section{Introduction}
An optical isolator is a device that allows the light to pass in one
direction but blocks it in the other. It is commonly used in laser technologies to shield the active medium of the lasers and amplifiers
from the back-reflected light's detrimental effect, which often leads to a number of light source instabilities.

\begin{figure}[h!]
\centerline{\includegraphics[width=8.2cm]{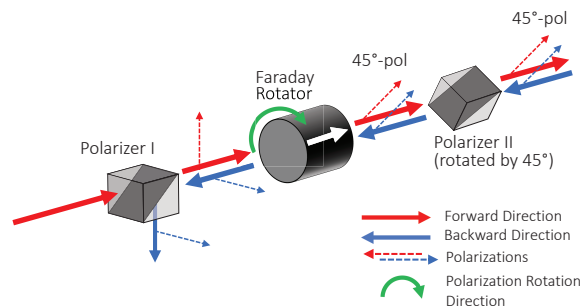}}
\caption{(Color online) Scheme of a Rayleigh-type optical isolator. Light propagating forward (red ray) is vertically polarized at polarizer I (a polarizing beam splitter here). The Faraday rotator leads to 45 degree linear polarization (red dashed arrows) which passes unaffected through polarizer II. Light
traveling backward (blue ray) is polarized at 45 deg by polarizer II and gets another 45 deg by the Faraday rotator (blue dashed arrows) so that it becomes  horizontally polarized. Consequently, the backward light is reflected or blocked by polarizer I.}
\label{fig1}
\end{figure}

The first optical isolator (OI) was suggested by Rayleigh \cite{Rayleigh1885} and consists of a Faraday rotator sandwiched between two polarizers as seen in Fig.~\ref{fig1}. If the polarization rotation at the  magneto-optical  Faraday rotator is 45 deg, the light passes unaffected through the second polarizer (with transmission axis oriented at 45 deg) and is therefore ideally transmitted in forward direction. In contrast, back-reflected light of the same polarization is getting blocked at polarizer I as a result of the non-reciprocal properties of the Faraday effect.
Using the above main principle, simple schema were realized by using different type of polarizers optimized for specific spectral regions, optical power levels, or for the pulsed rather than the continuous wave regime.

The major drawback of Rayleigh's optical isolators is to work in an optimum way only for one specific input light polarization that should be known in advance.  The same is true also for most recent magneto-optics-based isolators developed in integrated photonics, where either the TM or the TE polarization is generally being addressed (see \cite{Stadler2014,HuangOL17,Pintus,Srinivasan19,YanOptica20} and references therein).

Disposing of optical isolators operating independently of the state of the incoming light polarization is an attractive feature for both bulk optics and integrated implementations.
Different strategies have been developed to this aim.   One approach is based on using birefringent plates that induce spatial walk-off between the two orthogonal polarizations, what allows to treat them separately within  one or more non reciprocal Faraday elements  \cite{Nakajima,Shiraishi, Chang,ChangOL90}. Other types of  polarization independent isolators are based on Mach-Zehnder interferometers in waveguide configurations exhibiting magneto-optical nonreciprocal phase shifts for the TE and TM modes \cite{zhuromskyy99, FujitaPTL2000, Shoji}. Finally, various types of polarization independent circulators were developed either on a bulk optics \cite{Hidetoshi, Matsumoto, Emkey83,Shirasaki2, Yan, Fujii1, Fujii2, Koga92,Wang, Chen} or an integrated platform \cite{Sugimoto, Zaman, Firby18,Pintus}.

 Here we propose an alternative way to realize a polarization
independent optical isolator (PIOI) based on a common path interferometer and verify its proof of principle in a bulk optics arrangement.  The optical design is based on a Sagnac-type interferometer with the ring containing two nonreciprocal polarization switches (NRPS) put in series and intercalated by a polarizer. The NRPSs are composed of a Faraday rotator and a half-wave plate and leave the polarization unchanged for propagation in one direction but switches it between horizontal and vertical (rotation by 90 deg) in the opposite direction.  Importantly, in our PIOI the two   orthogonal polarizations are treated simultaneously.  They follow the same path within the ring but in counter propagating directions.  Section 2 presents the concept of the proposed PIOI, while Sect. 3 gives a proof-of principle experimental implementation. It is shown that the proposed PIOI conserves the input polarization to the output port and provides isolation exceeding 43 dB for any linear, circular or elliptical state of polarization.

\begin{figure}[t!]
\centerline{\includegraphics[width=8.2cm]{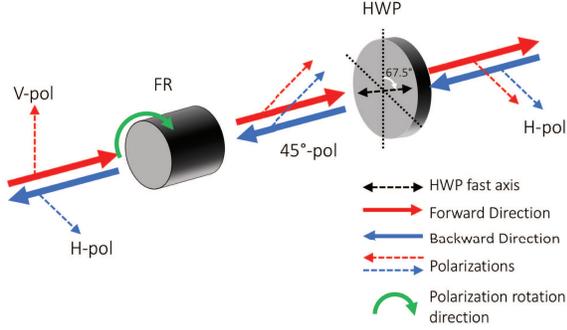}}
\caption{(Color online)  Principle of the nonreciprocal polarization switch (NRPS). FR is a Faraday rotator with 45 deg rotation angle and HWP is a half-wave plate oriented at 67.5 deg.
}
\label{fr+hwp}
\end{figure}

\section{Concept}
\begin{figure}[t!]
\centerline{\includegraphics[width=8.2cm]{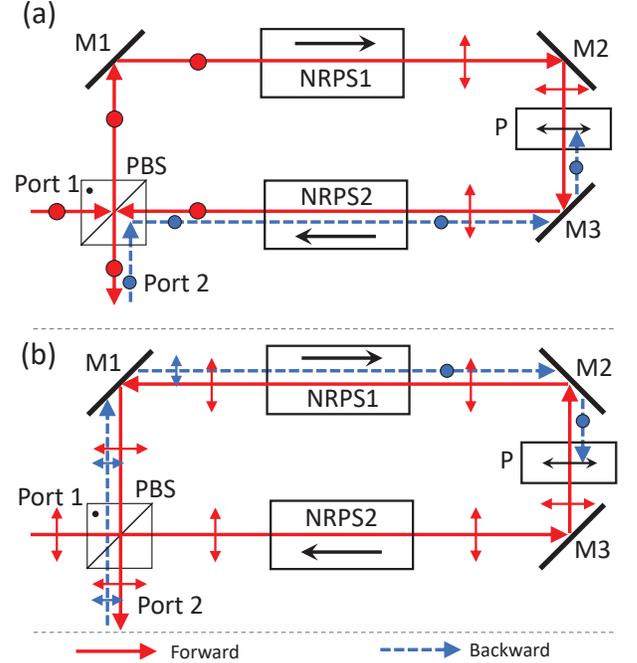}}
  \caption{(Color online) Polarization-independent optical isolator design (PIOI). In forward direction (red-solid arrows), the light travels from port-1 to port-2. In backward direction (blue-dashed arrows), the light traveling from port-2 towards port-1 is blocked at the horizontal polarizer (P). Frames (a) and (b) correspond to vertical and horizontal input polarizations, respectively.
PBS is a polarizing beam splitter, NRPS1 and NRPS2 are nonreciprocal polarization switches as of Fig.~\ref{fr+hwp}, and M1-M3 are mirrors.}
\label{PIOI}
\end{figure}

The principles underlying the NRPS and the PIOI are illustrated in Fig.~\ref{fr+hwp} and Fig.~\ref{PIOI}, respectively. The NRPS is composed of a Faraday rotator (FR) and a half-wave plate (HWP). The FR, as a non-reciprocal element, rotates the light polarization by 45 deg in the same direction for both light propagation directions and for all polarizations. In contrast, the HWP with one of its main axes oriented at 67.5 deg, rotates the polarization by +45 deg in the forward direction and by -45 deg in the backward one, provided that the input polarization is either horizontal or vertical. The overall effect of the NRPS is therefore to impose a 90 deg polarization rotation in forward direction (switch between the horizontal and vertical polarizations) and to leave the polarization unchanged in the backward one (see Fig.~\ref{fr+hwp}).  It is worth noting that the same NRPS functionality can be obtained also by replacing the HWP by a reciprocal optical rotator composed of a plate of an optically active crystal such as quartz, provided that the thickness is adjusted to achieve a rotation angle of 45 deg.

The PIOI working principle is illustrated for the specific cases of a vertical and horizontal input polarization in Fig.~\ref{PIOI}(a) and Fig.~\ref{PIOI}(b), respectively. Any light beam entering through the port 1 will exit the PIOI at port 2 with the same polarization as the input light. On the other hand, if the light re-enters the PIOI through port 2 in backward direction, due to the effect of the NRPSs the polarization is rotated in such a way that the whole wave is blocked by the horizontal polarizer (P) in the ring, so that no light can exit from port 1 in backward direction.
Specifically, the vertically polarized component of the input light (Fig.~\ref{PIOI}(a), red) is reflected by the polarizing beam splitter (PBS) and travels through the ring in clockwise direction passing both NRPSs in forward direction. NRPS1 changes its polarization to horizontal so that the light is fully transmitted through the horizontally aligned polarizer P. Further down NRPS2 switches its polarization back to vertical so that this wave component is reflected once more by the PBS  and is directed to the output port 2.
The reverse path is shown by the blue arrows in Fig.~\ref{PIOI}(a). If light with the same vertical polarization enters port 2 (backward PIOI direction) it gets reflected by the PBS but passes the element NRPS2 without change of its polarization and gets finally stopped by the horizontally aligned polarizer on its
way back towards port 1.

The case when the light travelling from port 1 to port 2 possesses
horizontal polarization is given in Fig.~\ref{PIOI}(b). This forward propagating light (red) is $p$-polarized at the PBS and is transmitted through this element, it travels through the ring in counter-clockwise direction with no change of polarization at any of the elements. When getting back to the PBS it is transmitted again towards port 2. For the reversed direction (Fig.~\ref{PIOI}(b), blue), a horizontally polarized beam entering the PIOI through port 2 follows the ring in clockwise direction and is blocked at the polarizer.

In summary, a forward input wave with arbitrary polarization is splitted by the PBS in its vertical polarization component (traveling clockwise through the ring) and its horizontal components (traveling counter-clockwise). Both will exit through port 2 after propagation through an equal path length in opposite directions. A backward wave from port 2 is also splitted by the PBS. However, in this case the vertical component travels counter-clockwise and the horizontal travels clockwise until reaching the polarizer and being blocked.

\section{Experiments}
Experimental proof of the proposed concept is realized by means of a bulk optics set-up corresponding to the design of Fig.~\ref{PIOI}.
All experiments were performed with a cw Ti:Sapphire laser adjusted to the wavelength  $\lambda=798$ nm. Both Faraday rotators (IO-3-780-HP, Optics for Research) in the NRPSs were also tuned to have 45 deg rotation at this wavelength. Two electrically tunable liquid crystals retarders (Thorlabs LCC1413- A) were calibrated to have the role of two HWP at this wavelength and their fast axes were rotated by 67.5 deg with respect to the vertical direction as shown in Fig.~\ref{fr+hwp}. The vertical and horizontal laboratory directions were defined by the positioning of the coated PBS at the entrance of the PIOI and were checked for all elements by means of light extinction through an auxiliary analyzer mounted on a rotation stage. The orientation of the Glan-Thompson crystal polarizer P in the PIOI ring was also adjusted in the same way.

\begin{figure}[t!]
\centerline{\includegraphics[width=8.2cm]{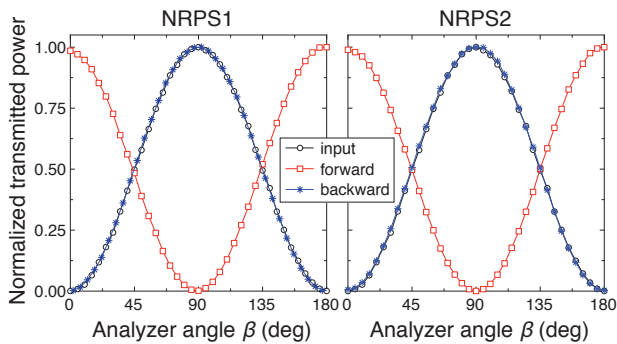}}
\caption{(Color online)
Characterization of the two NRPSs used in the experiments. The normalized transmitted power through an analyzer placed behind the NRPS is plotted as a function of the analyzer angle $\beta$, where $\beta=0$ gives light extinction in absence of the NRPS. The black circles are for the input light (without NRPS), the red squares are for NRPS1 or NRPS2 in forward direction (90 deg polarization rotation) and the blue stars are for the NRPSs in backward direction (no polarization rotation). }
\label{charact_NRPS}
\end{figure}

First we have verified the correct operation of both NRPSs and the results are shown in Fig.~\ref{charact_NRPS}. For this purpose we have prepared an input beam with horizontal polarization and measured the light extinction characteristics through an analyzer (Glan-Thompson polarizer) inserted after NRPS1 or NRPS2. The black curves in Fig.~\ref{charact_NRPS} are in absence of the NRPSs and give the normalized transmitted power through the analyzer as a function of its orientation angle $\beta$, with $\beta=0$ corresponding to the extinction position (vertical orientation). This situation gives the $\sin^2 \beta$ relationship expected from Malus law. The same is true for the blue curves that correspond to the case where NRPS1 (or NRPS2) is inserted in backward direction before the analyzer, what confirms that backward propagation through both NRPSs does not affect the polarization. In contrast, in the case where NRPS1 (or NRPS2) is inserted in forward direction (red curves) the normalized transmitted power is proportional to $\cos^2 \beta = \sin^2(\beta-\pi/2)$, confirming the 90 deg rotation in polarization.

\begin{figure}[t!]
\centerline{\includegraphics[width=8.2cm]{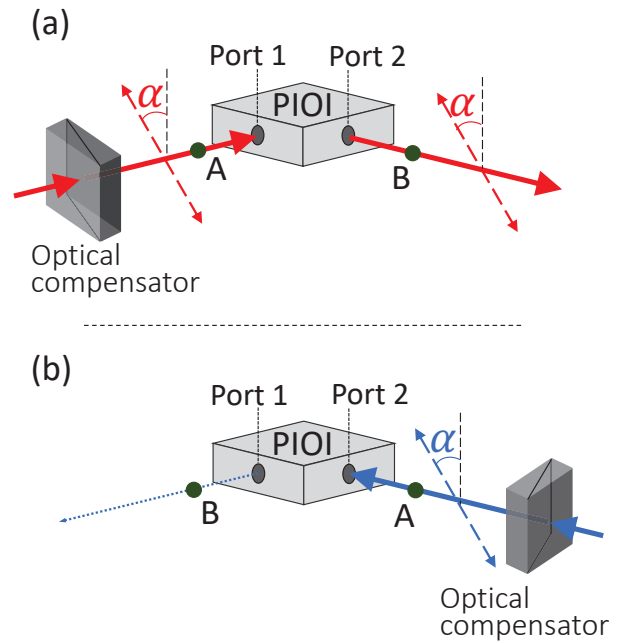}}
\caption{(Color online) Scheme for investigation of the PIOI in forward (a) and backward (b) directions. In both cases power detection occurs at points A and B, before the entrance port and after the exit port, respectively.  The optical compensator allows to prepare any input polarization state (here linearly polarized light under an angle $\alpha$ is shown). The PIOI is the ring arrangement in Fig.~\ref{PIOI}.
}
\label{fig4}
\end{figure}

The investigation of the polarization dependence and of the isolation characteristics of the PIOI were based on the scheme shown in Fig.~\ref{fig4}. We have inserted an additional  Soleil-Babinet optical compensator (a wave retarder with adjustable retardation)  before the PIOI, both when investigating the forward (Fig.~\ref{fig4}(a)) and the backward direction (Fig.~\ref{fig4}(b)). This allows to prepare different linear, circular or elliptical polarization states at the PIOI entrance. The points A and B shown on Fig.~\ref{fig4} correspond to the positions at which the powers are being detected. A combination of a non-polarizing beam splitter with known characteristics and photodetectors permits to determine the input power and the back-reflected power from the PIOI at point A. Similarly, a photodetector placed at point B leads to the determination of the transmitted power for the forward or backward direction.

\begin{figure}[t!]
\centerline{\includegraphics[width=8.2cm]{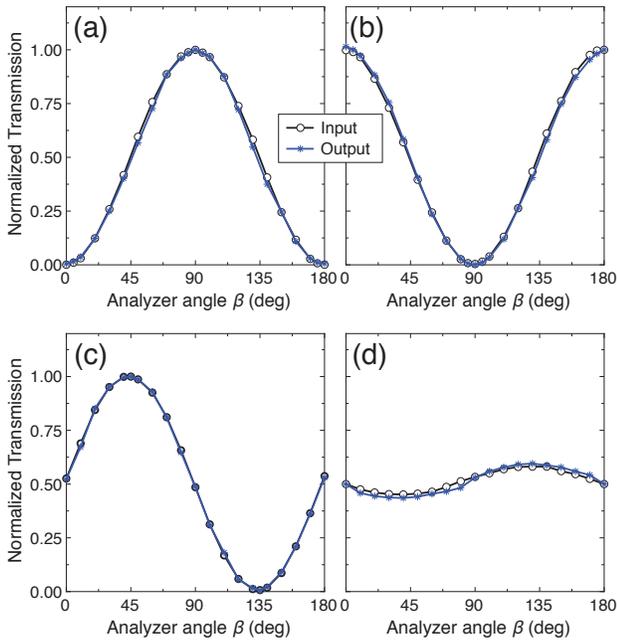}}
\caption{(Color online) Normalized transmission power through an analyzer for four different input polarization states: (a) horizontal, (b) vertical, (c) linear at nearly 45 deg and (d) elliptical. Measurement are for the forward direction through the PIOI.  The black circles are measured at the input (point A in Fig.~\ref{fig4}(a)) and the blue stars at the output (point B). The curves overlap confirms the conservation of the polarization state.
}
\label{forward_states}
\end{figure}

In order to prove that the PIOI conserves the input polarization we have proceeded in a way similar to the one used for the characterization of the NRPSs in Fig.~\ref{charact_NRPS}. An analyzer followed by a silicon photodetector were inserted alternatively at points A and B in Fig.~\ref{fig4}(a) and the normalized transmitted power upon rotation of the analyzer were taken in both cases. The results are shown in Fig.~\ref{forward_states} for four different states of input polarization: horizontal, vertical, linear at nearly (but not exactly) 45 deg and elliptical. The good overlap between the curves for the input and the output confirms the conservation of the polarization. Note that the direction of rotation of the analyzer angle $\beta$ at point B is opposite than at point A in order to take into account the inversion due to the reflections at an odd number of mirrors in the PIOI (three in our case). This is connected to the fact that, when always looking against the source, a wave linearly polarized under an angle $+\alpha$ will become polarized at $-\alpha$ upon reflection from one mirror or an odd number of mirrors, and a right circularly polarized wave will become left circularly polarized and vice-versa. This state of things can be recognized by the direction of the red polarization arrows in Fig.~\ref{fig4}(a). For the four states of polarization used for Fig.~\ref{forward_states} we also evaluated the back-reflection losses $\eta$ from the PIOI. The latter are defined as $\eta (dB)=10 \log_{10}\left[P_{input}/P_{back}\right]$, where $P_{input}$ and $P_{back}$ are the input and back-reflected power, respectively. For the four situations we obtained similar values, i.e. $\eta_h=41.9$ dB, $\eta_v=41.3$ dB, $\eta_{45} =41.9$ dB and $\eta_e=40.6$ dB for horizontal, vertical, 45 deg and elliptical polarized inputs, respectively.

For evaluating the isolation capability of the PIOI we have to compare the forward power transmission $T_{1\rightarrow2}$ from port 1 to port 2 with the reverse power transmission $T_{2\rightarrow1}$ in backward direction. The isolation level is then defined as
\begin{equation}
\zeta (dB)=10 \log_{10}\left[T_{1\rightarrow2}/T_{2\rightarrow1}\right] \ \ .
\label{eq:zeta}
\end{equation}

\begin{figure}[t!]
\centerline{\includegraphics[width=8.2cm]{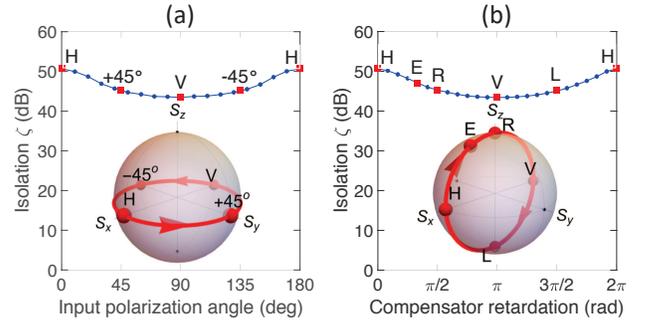}}
\caption{(Color online) Measured PIOI isolation $\zeta$ (dB) for various states of input polarization. In (a) all polarizations are linear and span the equator of the shown Poincar\'e sphere. In (b) they span a meridian line on the Poincar\'e sphere as obtained by varying the retardation of the input optical compensator. The labels H, V, $+45^\circ$, $-45^\circ$, E, R and L correspond to the horizontal, vertical, $+45^\circ$,$-45^\circ$, elliptical, right circular and left circular polarizations, respectively.
}
\label{linear+retard}
\end{figure}
Figure~\ref{linear+retard} gives the measured isolation for different states of polarization. In Fig.~\ref{linear+retard}(a) the polarization is linear and varies around the equator of the Poincar\'e sphere, as obtained by adjusting the compensator retardation to $\pi$ (HWP) and rotating this element around the beam propagation axis. For Fig.~\ref{linear+retard}(b), in contrast, the compensator has a fixed orientation at 45 deg but its retardation is varied between 0 and $2\pi$ at the used wavelength. This leads to a polarization variation along a meridian line on the Poincar\'e sphere, from horizontal (H) to various elliptical states (E), to right circular (R) and back to horizontal passing by the vertical (V) and the left circular state (L). As can be seen in Fig.~\ref{linear+retard} for all cases the isolation exceeds 43 dB and is larger than 50 dB for the horizontal polarization.

\begin{figure}[t!]
\centerline{\includegraphics[width=8.2cm]{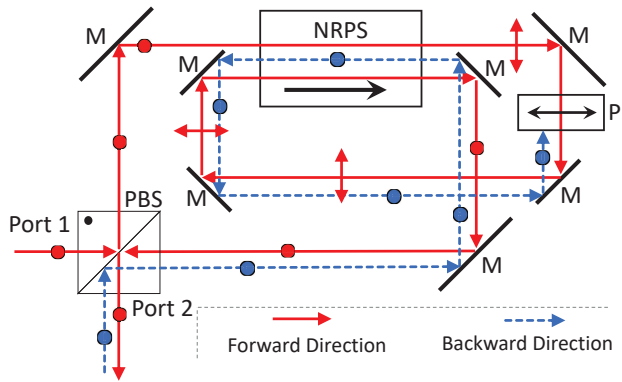}}
\caption{(Color online)   Folded configuration using a double pass in a single NRPS, corresponding to the case of Fig.~\ref{PIOI}(a) (vertical input polarization). M: mirrors, all other symbols as in Fig.~\ref{PIOI}. }
\label{folded}
\end{figure}

\section{Conclusion}
We have presented and verified experimentally an alternative concept for a polarization independent optical isolator. The fact that our scheme is based on a Sagnac-type ring interferometer brings about several advantages proper to common path interferometers. The most important one is the fact that there is no difference between the optical path of the two orthogonal polarizations, which allows to keep the polarization state between input and output. Also, such an arrangement features an increased robustness against length changes with respect to, for instance, Mach-Zehnder type isolators. Furthermore, the simple construction leads to easy tuning/adjustment. The implemented prototype was shown to conserve the input polarization state in forward direction and to provide an isolation exceeding 43 dB for all polarizations. We would like to mention that the use of two separated nonreciprocal polarization switches (and two Faraday rotators) is not strictly necessary. It is obviously possible to fold the optical path in such a way as to use the same NRPS twice  as depicted in Fig.~\ref{folded}.

\section*{Acknowledgment}

H. Hristova is supported by the Bulgarian Ministry of Education and
Science under the National Research Programme "Young scientists and
postdoctoral researchers" approved by DCM 577/17.08.2018. M. Al-Mahmoud is supported by the EU Horizon-2020 ITN project LIMQUET (Contract No. 765075). A. Rangelov is supported by Sofia University Grant 80-10-12/18.03.2021.

\end{document}